\documentstyle[11pt]{article}

\newcommand{\bigzerou}{%
\smash{\lower1.7ex\hbox{\bg 0}}}
\renewcommand{\theequation}{\arabic{section}.\arabic{equation}}
\newcommand{\beq}{\begin{equation}}
\newcommand{\enq}{\end{equation}}

\newcommand{\mapright}[1]{%
\smash{\mathop{%
\hbox to 1.0cm{\rightarrowfill}}\limits^{#1}}}
\newcommand{\mapleft}[1]{%
\smash{\mathop{%
\hbox to 1.3cm{\leftarrowfill}}\limits^{#1}}}

\newcommand{\no}{\nonumber}

\newtheorem{thm}{Theorem}

\newtheorem{defi}{Definition}
\newcommand{\beqy}{\begin{eqnarray}}
\newcommand{\enqy}{\end{eqnarray}}

\begin{document}


\begin{titlepage}
\vglue 3cm

\begin{center}
\vglue 0.5cm
{\Large\bf An Approach to ${\cal N}=4$ $ADE$ Gauge Theory
on $K3$}
\vglue 1cm
{\large Masao Jinzenji${}^\dagger $,  Toru Sasaki${}^* $} 
\vglue 0.5cm
{\it ${}^\dagger $
Division of Mathematics, Graduate School of Science,
        Hokkaido University, Sapporo 060-0810, Japan 
}\\
{\it ${}^*$Department of Physics,
Hokkaido University, Sapporo 060-0810, Japan}
{\it ${}^\dagger $jin@math.sci.hokudai.ac.jp \\
${}^*$ sasaki@particle.sci.hokudai.ac.jp}

\baselineskip=12pt

\vglue 1cm
\begin{abstract}
 We propose a recipe for determination of the partition function of 
${\cal N}=4$ $ADE$ gauge  theory on $K3$
by generalizing our previous results of the $SU(N)$ case.
The resulting partition function satisfies Montonen-Olive duality
for $ADE $ gauge group.
\end{abstract}
\end{center}
\end{titlepage}
\section{Introduction}
\label{sec:1}

This article is the continuation of our previous works \cite{jin,jin2}.
Based on the understanding obtained in these works, we try to determine 
the partition function of
${\cal N}=4$  $ADE$ gauge theory
on $K3$. 

${\cal N}=4$ theory is one of the good laboratories of 
testing duality conjecture.
It is well-known that
${\cal N}=4$ theory has the largest possible number of 
super-symmetry for a four-dimensional theory without gravity.
Furthermore,
it is believed to be exactly finite and conformally invariant.
The duality of this theory is believed to be the one proposed by Montonen 
and Olive. On the basis of this thought,
the study of ${\cal N}=4 $ $SU(N)$ gauge theory
has enjoyed many progresses with the aid of the $S$-duality conjecture:
the sharpened version of Montonen-Olive duality \cite{vafa-witten,lozano,m-v}.
On the other hand, there has been almost no attempt 
to determine the partition function
of $D,E$ gauge theory in spite of the fact that the $S$ duality conjecture 
for these cases has  already  been established \cite{vafa, M-O}. 

In our previous work, we have studied ${\cal N}=4$ super-symmetric
Yang-Mills theory on orbifold-$T^4/{\bf Z}_2$ for $SU(2)$ \cite{jin}
and for $SU(N)$ \cite{jin2}. In these works, we tried to realize 
the following geometrical picture at the partition function level.
A special class of $K3$ surfaces  
can be constructed from blowing up sixteen singularities of 
the quotient space $T^4/{\bf Z}_2=S_0$. 
Therefore, the partition function on $K3$ should be given as  
the product of the partition function of $S_0$
and the blow-up formula \cite{yoshioka,vafa-witten,kap}. 
This attempt was successful not only for $SU(2)$ but also for $SU(N)$.
The verification of this attempt mainly depends on
the following two facts.
The first one is that the $SU(N)$ partition function on $K3$
can be described by a Hecke transformation of order $N$
of $1/\eta(\tau)^{24}$\cite{vafa-witten,m-v,yoshihecke,lozano}.
The second one is that $1/\eta(\tau/N)$ can be written as the form
$\theta^2_{A_{N-1}}(\tau)/\eta(\tau)^N$,
where $\theta^2_{A_{N-1}}(\tau)$ is a kind of $A_{N-1}$ theta function
\cite{jin2}. This identity is easily verified from  
the celebrated  denominator identity of the affine Lie 
algebra \cite{kac,mac}.
These two facts enabled us to rewrite the partition function 
of $SU(N)$ theory on $K3$ into the desired form.

Looking back at the above verification, we can speculate that inversion 
of the above process can be used to guess the form of the partition 
function of general $ADE$ gauge theory.
Fortunately, we already have the works on $ADE$ blow-up formulas \cite{yoshioka,naka,kap}, which states that principal part of $ADE$ blow-up formulas have 
the form of $ADE$ theta functions $\theta_{A,D,E}(\tau)$. 
In this paper, we proceed further and assert that the partition function of 
$ADE$ gauge theory should take the form of $(\theta_{ADE}(\tau)/
\eta(\tau)^{r+1})^{24}$ ($r$ is the rank of the gauge group). This form can 
be verified by using the fact that ${\cal N}=4$ $ADE$ gauge theory on $K3$ is 
obtained from Type II A string theory compactified on $K3\times T^{2}\times 
(ALE\;\mbox{space})$.
By plugging this assertion  into the denominator identity,
we can see that the $D,E$ partition functions  can be rewritten  
as the form of $(\eta-\mbox{product})^{24}$.
As was pointed out before, 
the partition function of $SU(N)$ gauge theory on $K3$ surface 
is written only by $\eta$ function.  This is because the moduli
space of rank $N$ semi-stable sheaves is described by Hilbert scheme of 
$K3$ surface itself \cite{mukai,nak,yoshihecke}. Physical interpretation of 
this fact is also given from the point of view of Type IIA-Heterotic duality.
Though we have not obtain analogous results on the moduli space of $D, E$ 
vector bundles, we already have many beautiful results on $K3$ surface, 
which is enough for us to expect that the partition function of $D, E$ 
gauge theory should be written only by $\eta$-product. Therefore, we use 
the above $(\eta\mbox{-product})^{24}$ as the primary 
building block of the $D,E$ partition function on $K3$.
Next step of our construction is found from the analogy 
with  the $A_{N-1}$ case.
The corresponding primary building block of the $A_{N-1}$ partition function 
is $1/\eta(\tau/N)^{24} $. Then we take $SL(2,Z)$ orbit of this function 
(precisely speaking, we have to include the transformation 
$\tau \rightarrow \tau+1/m$ in some cases). After that, we take linear 
combination of all the building blocks so obtained, whose coefficients    
are determined to satisfy the $S$-duality conjecture, as was done  
in  Vafa-Witten 's original work \cite{vafa-witten}. The result is 
nothing but the order $N$ Hecke transformation of $1/\eta(\tau)^{24}$. 
Hence we applied the same procedure by taking $(\eta\mbox{-product})^{24}$
as the primary function.  As a result, we found that the above procedure 
also works quite well in this case.
Remarkably enough, the resulting partition function 
completely satisfies Montonen-Olive duality ! \cite{M-O}.

Unfortunately, our results are not complete from the geometrical point of 
view.
It is well-known that the partition function is
the generating function of the Euler number of
the moduli space of irreducible anti-self-dual connection.
For $ADE$ gauge group, we have Atiyah-Hitchin-Singer formula
that gives  us the dimension of the moduli space of irreducible  
anti-self-dual connection \cite{AHS}.
This formula tells us that if the instanton number $k$ is less 
than $\mbox{rank}({\cal G}_{ADE})+1$, the moduli space is empty.
On the other hand, the  partition function that we have derived    
has non-vanishing coefficients
in the region $k\le r$.
Up to now, we think that these phenomena is caused by existence of 
reducible connections.
Especially in the case of  $SU(N)$ theory on the manifold $b_2^+\le 1$, 
it is well-known that existence of reducible connections 
gives rise to the appearance of the holomorphic anomaly
in the partition function.
Thus we speculate that the true partition function of $D, E$ theory on $K3$
should have the  holomorphic anomaly.
In other words, we have derived the holomorphic part of
the partition function of $D,E$ gauge theory.
Therefore, we should search for an analogue of the holomorphic anomaly equation
used in \cite{m-v} as the second step. This problem will be discussed in our
subsequent paper.

The organization of this paper is the following:
In Sec.2 we review the Vafa-Witten theory 
in the $SU(N)$ case and discuss generalization to the $D,E$ gauge theory.
Next, we look carefully at the detailed structure of the duality conjecture 
on the partition function of $SU(N)$ theory as warming-ups.
In Sec.3 we estimate the form of the $D,E$ 
partition functions on $K3$ from the point of view of Type II A string theory 
compactified on $K3\times T^{2}\times (ALE\;\mbox{space})$.
In Sec.4 we combine the above result with the denominator identity of affine Lie algebra 
and  determine the $D,E$ partition function on $K3$
up to some small rank.
In Sec.5 we point out that our partition function of $D,E$ gauge theory on $K3$
does not satisfy the gap condition
and discuss the possible improvement parallel to the treatment of 
holomorphic anomaly in the $\frac{1}{2}K3$ case.
In Sec.6 we conclude and discuss the remaining problems.

\setcounter{equation}{0}
 
\section{Vafa-Witten Theory}
\label{sec:2}
\setcounter{equation}{0}

\subsection{'t Hooft Flux for ADE Gauge Theory}
In this subsection, we introduce the 't Hooft flux for
$ADE$ gauge theory. To this aim, we first consider the non-Abelian
version of the Dirac quantization condition \cite{M-O}:
\beq
\exp(4\pi ieF_0)=1,\label{dqc}
\enq 
where $F_0$ is a Lie algebra valued matrix of $ADE$ gauge group ${\cal G}$
and appears in the solution of the magnetic
monopole equation. By using gauge transformation, $F_0$ can be rewritten into 
the form:
\beq
eF_0=\sum_{j=1}^{rank({\cal G})}\beta_j T_j,
\enq
where the $T_j$'s are any appropriately normalized set of mutually
commutative generators of ${\cal G}$. Here we introduce the root lattice
$\Lambda ({\cal G}) $. Then (\ref{dqc}) turns out to be
\beq
2\beta\cdot\alpha \in {\bf Z},~~\mbox{for all}~~\alpha\in \Lambda ({\cal G}). 
\enq
This condition tells us that $\beta$ should belong to the dual 
lattice (or weight lattice)
$\Lambda ({\cal G}^\vee) $. On the other hand,
it is well-known that 
\beq
\Lambda ({\cal G}^\vee)/\Lambda ({\cal G})\cong \Gamma_{\cal G},
\enq
where $\Gamma_{\cal G}$ is center of ${\cal G}$.
In this way, Montonen and Olive proposed that $ADE$ gauge group ${\cal G}$
is dual to ${\cal G}/\Gamma_{\cal G}$.
From this proposal,
it is natural to assume that 't Hooft flux of the dual group
${\cal G}/\Gamma_{\cal G}$ is
given by $v\in H^2(X,\Gamma_{\cal G})$. For later use, we write down     
$ \Gamma_{\cal G}$ in the following table:
\begin{table}[h]
\begin{center}
\begin{tabular}{|c|c|c|}
\hline
${\cal G}$& $\Gamma_{\cal G}$\\
\hline
$A_{N-1}$ & ${\bf Z}_N$ \\
\hline
$D_{2N}$ &  ${\bf Z}_2\times{\bf Z}_2$\\
\hline
$D_{2N+1}$ &  ${\bf Z}_4$\\
\hline
$E_6$ &  ${\bf Z}_3$\\
\hline
$E_7$ &  ${\bf Z}_2$\\
\hline
$E_8$ &  trivial\\
\hline
\end{tabular}
\end{center}
\end{table}
\\
Classification of the type of the {}'t Hooft flux of 
$A_{N-1}$ ($N$: prime) is already studied \cite{vafa-witten,lozano}.
However, general $N$ case is more complicated.
In Sec.2.3, we will show explicitly how $v\in H^2(K3,\Gamma_{\cal G})$
are classified in the case of general $ADE$ gauge group.

\subsection{Vafa-Witten Conjecture for ADE Gauge Theory}
Following \cite{vafa-witten,lozano,yoshi}, we review the
general structure of Vafa-Witten conjecture.
Twisted ${\cal N}=4$ gauge theory on the manifold 
with suitable vanishing theorem
causes remarkable simplification \cite{vafa-witten}.
That is, its partition function has the form of 
the summation of the Euler number of the moduli space of
the ASD equations. 
More precisely,
for twisted ${\cal N}=4$ ${\cal G}/\Gamma_{\cal G}$ gauge 
theory with 't Hooft flux $v\in H^{2}(X,\Gamma_{\cal G})$ on $X$,
we propose that the partition function of this theory is 
given by the formula: 
\begin{equation}
Z^X_v(\tau):= q^{-{\frac{(r+1)\chi(X)}{24}}}\sum_k \chi({\cal M}(v,k))q^k
\;\;\;(q:=\exp(2\pi i \tau)),\label{zxvt}
\end{equation}
where ${\cal M}(v,k)$ is 
the moduli space of ASD connections
associated to ${\cal G}/\Gamma_{\cal G}$-principal bundle with
{}'t Hooft flux $v$ 
and fractional instanton number 
$k\in \frac{1}{2\vert \Gamma_{\cal G}\vert}{\bf Z}$.
In (\ref{zxvt}), $\tau$ is the gauge coupling constant 
including theta angle, 
$\chi(X)$ is Euler number of $X$ and $r$ is the rank of ${\cal G}$.
Introduction of $q^{-\frac{1}{24}} $ factor
is required by the modular property like the case of $\eta(\tau) $ function.
For $SU(N)/{\bf Z}_N$, this conjecture is nothing but Vafa-Witten's 
\cite{vafa-witten}.
With this result, Vafa and  Witten conjectured the behavior 
of the partition functions under the action of $SL(2,{\bf Z})$ on 
$\tau$.
They started with 't Hooft's work \cite{tHooft} in mind.
In \cite{tHooft}, the path integral with ${\bf Z}_N $-valued electric flux
and that with magnetic flux 
is related by Fourier transform.
Vafa and Witten combined the conjecture of strong/week duality
to this 't Hooft's result. 
Their conjecture is summarized by the following formula:
\begin{equation}
Z^X_v\left(-\frac{1}{\tau}\right)=\vert\Gamma_{\cal G}\vert^{-\frac{b_2(X)}{2}}
\left(
\frac{\tau}{i}
\right)^{-\frac{\chi(X)}{2}}
\cdot
\sum_{u\in H^2(X,\Gamma_{\cal G})}
\zeta_{\cal G}^{u\cdot v}Z^X_u(\tau),
\label{vw}
\end{equation}
where $\zeta_{\cal G}=\exp(\frac{2\pi i}{N_{\cal G}}) $.
We conjecture that this duality relation is also valid for the general 
$ADE$ partition function (\ref{zxvt}).

For later use, we introduce the notation:
\[
Z_{\cal G}^X(\tau):= \frac{1}{\vert\Gamma_{\cal G}\vert} Z_t^X(\tau),
\]
\begin{equation}
Z_{{\cal G}/\Gamma_{\cal G}}^X(\tau):= \sum_{u\in H^2(X,\Gamma_{\cal G})} Z_u^X(\tau).
\end{equation}
In this notation, we can obtain the following 
formula from (\ref{vw}):
\begin{equation}
Z^X_t\left(-\frac{1}{\tau}\right)=\vert\Gamma_{\cal G}\vert^{-\frac{b_2(X)}{2}}
\left(
\frac{\tau}{i}
\right)^{-\frac{\chi(X)}{2}}
Z_{{\cal G}/\Gamma_{\cal G}}^X(\tau).
\label{m-o}
\end{equation}
This formula is one of the key points on their explicit determination 
of the form of the partition function of complex surface with ample 
canonical bundle.

\subsection{Counting the Number of Orbits on $K3$}
In the rest of this paper, we will concentrate on $ADE$ gauge theory on $K3$.\\
In \cite{vafa-witten}, Vafa and Witten classified the types of 
$SO(3)=SU(2)/{\bf Z}_2$ 
theory with 't Hooft flux $v\in H^2(K3,{\bf Z}_2)$ on $K3$. 
Note that $b^2(K3)=22$, and $v$ can takes $2^{22}$ values.
They used the property that
diffeomorphism invariant of $v$ is given by $v^2$ mod 4 ($v\neq 0$) and $v=0$.
Especially on $K3$ there are only three types of partition functions.
These three types are called as even ($v^2\equiv 0$ mod $4$ and $v\ne 0$),
odd ($v^2\equiv 2$ mod $4$) and trivial ($v=0$) respectively.
Furthermore numbers of each type are denoted as $n_{even},n_{odd}$ and $n_0$
respectively. These numbers are concretely counted in \cite{vafa-witten}.
Of course they sum up to $n_0+n_{even}+n_{odd}=2^{22}$.
It is well known that the intersection form on
$H^2(K3,{\bf Z})$ is $H^{\oplus 3}\oplus (-E_8)^{\oplus 2}$.
However Vafa and Witten took  
$H^{\oplus 11}$ (the direct sum of 11 copies of $H$) 
instead of 
$H^{\oplus 3}\oplus(-E_8)^{\oplus 2}$ itself
as the intersection form on
$H^2(K3,{\bf Z})$ for convenience.
Here 
\beq
H=\left(\begin{array}{cc}
0&1\\
1&0
\end{array}\right).
\enq
This replacement makes counting the numbers of orbits very simple.
We follow this replacement in this paper.
From now on, we generalize the above result of the $SU(2)$ case to the case of 
general $ADE$ group.
First we count  naive number of orbits for $SU(N)/{\bf Z}_N$.
We denote  't Hooft flux for $SU(N)/{\bf Z}_N$ by,
\beq  
v_N\in (H^{\oplus 11})\otimes_{\bf Z}{\bf Z}_{N}=:H_{N}.
\enq
Roughly speaking, the word ``naive'' means that $v_N$ has no restriction.
Explicitly, the naive number of orbits of 
$SU(N)/{\bf Z}_N$ is defined as follows:
\begin{eqnarray}
n_j(N)&=&
\mbox{number of }v_N^2\equiv 2j \mbox{ mod }2N.
\end{eqnarray}
We introduce $V$ whose intersection form 
is given by $H$. For $V$, we define, 
\beq
l_j=\mbox{numbers of~}V^2\equiv 2j \mbox{~mod~} 2N.
\enq
With this definition, one can build the recursive formula
for computing the number of orbits on the manifold with $H^{\oplus n}$.
The solution for $n=11$ case is 
\begin{eqnarray}
n_j(N)&=&\frac{1}{N}\sum_{k=0}^{N-1}m_k^{11}\zeta_N^{jk},
\end{eqnarray}
where
\beq
m_k=\sum_{j=0}^{N-1}l_j\zeta_N^{jk},
\enq
and
\[
\zeta_N=\exp(\frac{2\pi i}{N}).
\]
From now on, we introduce a subscript "$t$" such as $n_t$,
which corresponds to the trivial flux $v=0$.
\paragraph{$A_{N-1}$: $N$ prime}~\\
In this part, we count the number of orbits of $SU(N)/{\bf Z}_N$:$N$ prime.
If $N$ is prime number, one  finds $m_0=2N-1,m_j=N,(j=1,\ldots,N-1)$.
With some algebra, one can find 
\begin{eqnarray}
n_0^N&=&
\mbox{number of }v_N^2\equiv 0 \mbox{ mod }2N \mbox{ and }v_N\ne 0
\no
\\ 
&=&
n_0(N)-1=
N^{21}+(N-1)N^{10}-1,\\
n_j^N&=&
\mbox{number of }v_N^2\equiv 2j \mbox{ mod }2N 
\no
\\ 
&=&
n_j(N)=
N^{21}-N^{10},j=1,\ldots,N-1,\\
n_t^1&=&
\mbox{number of }v_N=0 
\no
\\ 
&=&
n_t=1.
\end{eqnarray}
\paragraph{$A_{N-1}$: $N$ non-prime}~\\
In this part, we count the number of orbits of $SU(N)/{\bf Z}_N$:$N$ non-prime.
That is $N=p_1^{m_1}p_2^{m_2}\cdots p_l^{m_l}$ case ($p_j$ is a prime).
Here we took the following two cases as typical examples:
(i) $N=p^m$, (ii) $N=pq$.
Other cases are straightforward generalization of these two cases.
Before moving to specific discussion, 
we explain general structure of
non-prime $N$ case. 
For non-prime $N$, there are several blocks of type, 
which are classified by all possible divisor of 
$N=p_1^{m_1}p_2^{m_2}\cdots p_l^{m_l}$.
That is
\[
p_1^{M_1}p_2^{M_2}\cdots p_l^{M_l},0\le M_1\le m_1,
0\le M_2\le m_2,\ldots,0\le M_l\le m_l.
\]
Each block labeled by $p_1^{M_1}p_2^{M_2}\cdots p_l^{M_l}$ 
has the types $j\;\;(0\leq j\leq 
p_1^{M_1}p_2^{M_2}\cdots p_l^{M_l}-1)$.
Thus we denote 
the number of orbits of block $m$ and type $j$
by $n^m_j$.
Now, we give explicit definitions  of these materials.
First, we define $C_{p_1^{M_1}p_2^{M_2}\cdots p_l^{M_l}}\subset H_{N}$  
as follows.
\begin{equation}
C_{p_1^{M_1}p_2^{M_2}\cdots p_l^{M_l}}:=
\{v_{N}\in H_{N}\;|\;p_1^{M_1}p_2^{M_2}\cdots p_l^{M_l}v_{N}=0\}
\label{cs}
\end{equation}
Notice that $C_{m}\subset C_{n}$ if and only if $m|n$, and we can easily see 
the following equality:
\begin{equation}
C_{i_{1}}\cap C_{i_{2}}\cap C_{i_{3}}\cap\cdots \cap C_{i_{j}}
=C_{\mbox{gcd}(i_{1},i_{2},\cdots,i_{j})}.
\end{equation} 
Since the condition $p_1^{M_1}p_2^{M_2}\cdots p_l^{M_l}v_{N}=0$ means that 
all the components of $v_{N}$ must be multipliers of $p_1^{m_{1}-M_1}p_2^{m_{2}-M_2}\cdots p_l^{m_{l}-M_l}$, we can construct a natural isomorphism:
\begin{eqnarray}
&&\phi_{p_1^{M_1}p_2^{M_2}\cdots p_l^{M_l}}:
C_{p_1^{M_1}p_2^{M_2}\cdots p_l^{M_l}}\rightarrow 
H_{p_1^{M_1}p_2^{M_2}\cdots p_l^{M_l}},\no\\ 
&&\phi(v_{N}):=
\frac{1}{p_1^{m_{1}-M_1}p_2^{m_{2}-M_2}\cdots p_l^{m_{l}-M_l}}v_{N}\in 
H_{p_1^{M_1}p_2^{M_2}\cdots p_l^{M_l}}.
\end{eqnarray}
With these set-up, we define the block 
$B_{p_1^{M_1}p_2^{M_2}\cdots p_l^{M_l}}$ as follows:
\begin{equation}
B_{p_1^{M_1}p_2^{M_2}\cdots p_l^{M_l}}:=
C_{p_1^{M_1}p_2^{M_2}\cdots p_l^{M_l}}\setminus
\biggl(\cup_{m|p_1^{M_1}p_2^{M_2}\cdots p_l^{M_l},\;\;m\neq 
p_1^{M_1}p_2^{M_2}\cdots p_l^{M_l}}(C_{m})\biggr).
\end{equation}
Then the number $n_{j}^{p_1^{M_1}p_2^{M_2}\cdots p_l^{M_l}}$ is nothing but
the number of $v_{N}\in B_{p_1^{M_1}p_2^{M_2}\cdots p_l^{M_l}}$ 
that satisfies the condition:
\begin{equation}
(\phi_{p_1^{M_1}p_2^{M_2}\cdots p_l^{M_l}}(v_{N}))^{2}\equiv
2j \mbox{~mod~} 2p_1^{M_1}p_2^{M_2}\cdots p_l^{M_l}.
\end{equation} 
\\
(i) $N=p^m$ case\\
In this case, we have $m$ blocks $B_{p^{M}}=C_{p^{M}}\setminus C_{p^{M-1}}
\;(M=0,1,2,\cdots,m)$ and each block $B_{p^{M}}$ has $p^{M}$ types.
Notice that we have the following inclusion sequence in this case:
\begin{equation}
\{0\}=C_{1}\subset C_{p}\subset C_{p^{2}}\subset \cdots \subset C_{p^{m-1}}
\subset C_{p^{m}}=H_{p^{m}}.
\end{equation} 
Since we have the isomorphism $\phi_{p^{M}}: C_{p^{M}}\rightarrow H_{p^{M}}$
, we can write down $n^{p^{M}}_{j}$ in terms of naive numbers $n_{i}(p^{M})$ 
and $n_{l}(p^{M-1})$. The result is the following: 
\\
Block $B_{p^{M}},(M=m,m-1,\ldots,1)$
\begin{eqnarray}
n^{p^M}_{kp^{2}}&=&
n_{kp^2}(p^M)-\sum_{l=0}^{p-1}n_{lp^{M-2}+k}(p^{M-1}),k=0,\ldots,p^{M-2}-1,
\end{eqnarray}
\beqy
n^{p^M}_j&=&
n_j(p^M),j=1,\ldots,p^M-1 \mbox{ and } j\ne kp^{M-2},
\enqy
Block $B_{1}$,
\beqy
n_t^1&=&n_t=1. 
\enqy
For later use,
we explicitly write down the results in the $N=4$ case.\\ 
$N=4$ case
\begin{eqnarray}
n_0^{4}&=&n_0(4)-n_0(2)-n_1(2)=2^{42}+2^{31}-2^{21},\no\\
n_1^{4}&=&n_1(4)=2^{42}-2^{31},\no\\
n_2^{4}&=&n_2(4)=2^{42}+2^{31}-2^{21},\no\\
n_3^{4}&=&n_3(4)=2^{42}-2^{31},\no\\
n_0^{2}&=&n_0(2)-n_t=2^{21}+2^{10}-1,\no\\
n_1^{2}&=&n_1(2)=2^{21}-2^{10}\no,\\
n_t^1&=&n_t=1.
\end{eqnarray}
(ii) $N=pq$ case\\
In this case, we have 4 blocks:
\begin{equation}
B_{pq}=C_{pq}\setminus(C_{p}\cup C_{q}),\;B_{p}=C_{p}\setminus C_{1},
B_{q}=C_{q}\setminus C_{1},\;B_{1}=C_{1}=\{0\},\;\; 
(C_{p}\cap C_{q}=C_{1}). 
\end{equation}
Therefore, we can evaluate  $n^{m}_{j}$ by use of naive numbers 
in the same way as the previous case. 
\\
Block  $B_{pq}$,
\beqy
n^{pq}_0&=&n_{0}(pq)-n_0(q)-n_0(p)+n_t,
\enqy
\beqy
n_{kp}^{pq}&=&n_{kp}(pq)-n_{l^q_k}(q),\;\;k=0,\ldots,q-1,
\enqy
\beqy
n_{kq}^{pq}&=&n_{kq}(pq)-n_{l^p_k}(p),\;\;k=0,\ldots,p-1,
\enqy
\beqy
n_j^{pq}&=&n_j(pq),\;\;j=1,\ldots,pq-1,\;\;j\ne kp,\;\;j\ne kq, 
\enqy
where $l^q_k$ and $l^p_k$ satisfy  
\beq
l^q_kp^2\equiv 2kp \mbox{ mod }2pq,\;\;
l^p_kq^2\equiv 2kq \mbox{ mod }2pq.
\enq
Block  $B_p$,
\beqy
n_0^{p}&=&n_0(p)-n_t,
\enqy
\beqy
n_j^{p}&=&n_j(p),\;\;j=1,\ldots,q-1,
\enqy
Block $B_q$, numbers are given by $p\leftrightarrow q$.\\
Block $B_1$,
\beqy
n_t^1&=&n_t=1. 
\enqy
In the rest of this subsection, we count the number of orbits of 
${\cal G}/\Gamma_{\cal G}$.
However the counting is the same as $SU(N)/{\bf Z}_N$ case 
except for $D_{2N}/{\bf Z}_2\times{\bf Z}_2$. 

\paragraph{$D_{2N+1}/{\bf Z_4}$,~$E_6/{\bf Z}_3 $ and $E_7/{\bf Z}_2$}~\\
The counting of $D_{2N+1}/{\bf Z_4}$,~$E_6/{\bf Z}_3 $ and $E_7/{\bf Z}_2$
cases is  the same as those of $SU(4)/{\bf Z}_4$, $SU(3)/{\bf Z}_3$ and 
$SU(2)/{\bf Z}_2$
cases respectively, because our counting only depends on the center 
$\Gamma_{\cal G}$.

\paragraph{$E_8 $}~\\
$E_8$ has only one orbit and the result is
\begin{eqnarray}
n_t^1&=&1.
\end{eqnarray}
This result is just a reflection of the well-known fact that $E_{8}$ 
lattice is self-dual.

\paragraph{$D_{2N}/{\bf Z}_2\times{\bf Z}_2$}~\\
$D_{2N}/{\bf Z}_2\times{\bf Z}_2$ must be classified by
$v^2$ mod $2$, because $v^2$ takes value on ${\bf Z}_2$.
This classification produces degeneration 
${\bf Z}_2\times{\bf Z}_2\to {\bf Z}_2\;\;((m,n)\to m+n)$.
The correct counting including this degeneration
is given by the following treatment.
We regard the counting problem in this case as the one  
of $SU(2)/{\bf Z}_2$ theory
on a manifold with the replaced intersection form
$H^{\oplus 11}\to H^{\oplus 22}$.
This treatment reproduces the desired degeneration
and the correct numbers. The result is

\begin{eqnarray}
n_0^2&=&2^{43}+2^{21}-1,\no\\
n_1^2&=&2^{43}-2^{21},\no\\
n_t^1&=&1.
\end{eqnarray}

\subsection{Partition Function for $SU(N)$ Gauge Theory on $K3$}
Following \cite{vafa-witten,lozano,m-v}, we derive the $SU(N)$ partition
function on $K3$ as a reconstruction of \cite{vafa-witten,lozano,m-v}.
By using the picture that ${\cal N}=4 ~U(1)$ gauge theory
is given by single $M5$ brane wrapped around $K3\times T^2$,
one can understand that
$U(1)$ partition function on $K3$ is given by 
\beq
\frac{1}{\eta(\tau)^{24}}=:G(\tau).
\enq
Here we ignore the contribution from  the $U(1)$ fluxes.
 
To obtain ${\cal N}=4 ~U(N)$ gauge theory on $K3$,
we consider $N$ coincident $M5$ brane wrapped around $K3\times T^2$.
Furthermore by fixing 't Hooft flux trivial $v=0$,
we obtain ${\cal N}=4 ~SU(N)$ gauge theory on $K3$.
In this picture, Hecke transformation naturally appears as the effect of 
summing up all the ways of wrapping $T^{2}$ in $K3\times T^{2}$ $N$ times 
by $T^{2}$ in $M5$ brane.   
However in the remaining subsection,
we apply a more mechanical method 
for considering the generalization to $D,E$
gauge theory. 
Of course the result is the same as the result from the above picture.

First we consider the primary function
\beq
G_0(\tau):=\left(\frac{1}{\eta(\frac{\tau}{N})}\right)^{24},
\enq
as a piece of $Z_{SU(N)}(\tau)$. Next, by transforming $G_0(\tau)$
by $SL(2,{\bf Z})$, we obtain the other pieces of $Z_{SU(N)}(\tau)$.
Following \cite{vafa-witten}, we introduce
some combination of the above functions $Z_j(\tau)$ corresponding to 
the $SU(N)/{\bf Z}_N$ partition function with $v_N^2\equiv 2j$ mod $2N$. 
By requiring Montonen-Olive duality (\ref{m-o}) 
between $Z_t(\tau)$ and $Z_{SU(N)/{\bf Z}_N}$,
we can determine the coefficients of $G$'s in $Z_j(\tau)$.\\ 
Let us move to writing out the concrete $SU(N)$ partition function  
for three cases, prime $N$, non-prime $N=p^m$ and $N=pq$. 

\paragraph{$A_{N-1} $: $N$ prime}
By transforming $G_0(\tau)$ by $SL(2,{\bf Z})$, we obtain $N+1$ functions,
\beq
G_j(\tau):=\left(
\frac{1}{\eta(\frac{\tau+j}{N})}
\right)^{24},j=0,\ldots,N-1,
\enq
\beq
H_0(\tau):=\left(
\frac{1}{\eta(N\tau)}
\right)^{24}.
\enq
Following \cite{vafa-witten}, we define the following partition functions,
\beq
Z_j(\tau):=\frac{1}{N}\left(
G_0(\tau)+\zeta_N^{-j}G_1(\tau)+\cdots +\zeta_N^{-j(N-1)}G_{N-1}(\tau)
\right),j=0,...,N-1,
\enq
\beq
Z_t(\tau):=\frac{1}{N^2}H_0(\tau)+Z_0(\tau),\label{zpt1}
\enq
From these partition functions and the numbers in Sec.2.3, we obtain
\beq
Z_{SU(N)/{\bf Z}_N}(\tau):=\sum_v Z_v(\tau)=
n_0Z_0(\tau)+\cdots+n_{N-1}Z_{N-1}(\tau)+n_tZ_t(\tau).\label{zpt2}
\enq
(\ref{zpt1}) and (\ref{zpt2}) satisfy the Montonen-Olive duality (\ref{vw}).
We will call the functions appeared in the partition functions as $G$ 
functions. Most of the  $G$ functions are obtained from $SL(2,{\bf Z})$ 
transformation of $G_0(\tau)$, but some $G$ functions must be added such 
as in the following non-prime $N$ cases.

Determination of the partition function for non-prime $N$ case 
has the same structure as that of 
the counting the number of orbits. The result is the following.
\paragraph{$A_{N-1} $: $N$ non-prime}~\\
(i)$N=p^m$ case\\
We introduce the following $G$ functions of block $p^M$ and type $j$:\\
Block $p^M,(M=m,m-1,\ldots,0)$,
\beq
G_j^{p^M}(\tau):=\left(
\frac{1}{\eta(\frac{p^{m-M}\tau+j}{p^M})}
\right)^{24},j=0,\ldots,p^{M}-1.
\enq
Note that most of these functions are obtained by 
$SL(2,{\bf Z})$ transformation
from $G^{p^m}_0(\tau)$. However we have to introduce some $G$ functions
(which are not obtained by $SL(2,{\bf Z})$ transformation
from $G^{p^m}_0(\tau)$ ), so that we obtain the set of partition functions 
corresponding to all the types considered in Sec.2.3.
\\
By using $G^{p^m}_j(\tau), (j=0,\ldots,p^m-1)$, we introduce the partition function
of block $p^m$ and type $j$:  
\beq
Z_j^{p^m}(\tau):=\frac{1}{p^m}\left(
G_0^{p^m}(\tau)+\zeta_{p^m}^{-j}G_1^{p^m}(\tau)+\cdots +\zeta_{p^m}^{-j(p^m-1)}G_{p^m-1}^{p^m}(\tau)
\right),j=0,\ldots,p^m-1.
\enq
For the partition function
of block $p^M,(M=m-1,\ldots,1)$ and type $j$, we introduce
\beqy
Z_{lp^{M-1}+k}^{p^{M}}(\tau)&:=&\frac{1}{p^{2m-M}}\left(
G_0^{p^{M}}(\tau)+\zeta_{p^{M}}^{-(lp^{M-1}+k)}G_1^{p^{M}}(\tau)+\cdots 
+\zeta_{p^{M}}^{-(lp^{M-1}+k)(p^{M}-1)}G_{p^{M}-1}^{p^{M}}(\tau)
\right)
\no\\
&&
+Z^{p^{M+1}}_{kp^2}(\tau)
,\;\;l=0,\ldots,p-1,\;\;k=0,\ldots,p^{M-1}-1.
\enqy
Note that the partition function $Z_{lp^{M-1}+k}^{p^{M}}(\tau) $
consists of $G^{p^M}_j(\tau), (M=0,\ldots,p^M-1) $, and the partition function 
$Z^{p^{M+1}}_{kp^2}(\tau) $. This is the same reason why
the number of orbits of block $p^l$ 
is calculated by the naive number of orbits of block $p^l$ and
that of block $p^{l-1}$.\\
Finally for the partition function of block $1$, we introduce
\beqy
Z_t^1(\tau)&:=&
\frac{1}{p^{2m}}G^1_0(\tau)+Z_0^{p}(\tau)
\no\\&=&\frac{1}{p^{2m}}\sum_{0\le a,b,d \in{\bf Z} \atop ad=p^m,b<d}
dG(\frac{a\tau+b}{d}).
\label{zpmt}
\enqy
In the same way as the prime $N$ case, we obtain
\beqy
Z_{SU(p^m)/{\bf Z}_{p^m}}(\tau)&:=&
\sum_{v_N}Z_{v_N}(\tau)
\no\\&=&
n_0^{p^m}Z_0^{p^m}(\tau)+\cdots+n_{p^m-1}^{p^m}Z_{p^m-1}^{p^m}(\tau)
\no\\&&
+n_0^{p^{m-1}}Z_0^{p^{m-1}}(\tau)+\cdots
+n_{p^{m-1}-1}^{p^{m-1}}Z_{p^{m-1}-1}^{p^{m-1}}(\tau)
\no\\&&
+\cdots+n_0^{p}Z_0^{p}(\tau)+\cdots+n_{p-1}^{p}Z_{p-1}^{p}(\tau)
+n_t^1Z_t^1(\tau)
\no\\
&=&\frac{1}{p^{2m}}\sum_{0\le a,b,d \in{\bf Z} \atop ad=p^m,b<d}
d^{12}(\mbox{gcd}(b,d))^{11}G(\frac{a\tau+b}{d}).
\label{zpmd}
\enqy
Last equalities in (\ref{zpmt}) and (\ref{zpmd}) are nothing but 
the result in \cite{lozano}.\\
(ii)$N=pq$ case\\
We introduce the following $G$ functions by $SL(2,{\bf Z})$ transformation
from $G^{pq}_0(\tau)$,\\
\beq
G_j^{pq}(\tau):=\left(
\frac{1}{\eta(\frac{\tau+j}{pq})}
\right)^{24},j=0,\ldots,pq-1.
\enq
Block $p$,
\beq
G_j^{p}(\tau):=\left(
\frac{1}{\eta(\frac{q\tau+j}{p})}
\right)^{24},j=0,\ldots,p-1,
\enq
Block $q$,
\beq
G_j^{q}(\tau):=\left(
\frac{1}{\eta(\frac{p\tau+j}{q})}
\right)^{24},j=0,\ldots,q-1,
\enq
Block $1$,
\beq
G_t^{1}(\tau):=\left(
\frac{1}{\eta(pq\tau)}
\right)^{24}.
\enq
For the partition function of block $pq$, we introduce
\beq
Z_{j}^{pq}(\tau):=\frac{1}{pq}\left(
G_0^{pq}(\tau)+\zeta_{pq}^{-j}G_1^{pq}(\tau)+\cdots +\zeta_{pq}^{-j(pq-1)}G_{pq-1}^{pq}(\tau)
\right),j=0,\ldots,pq-1,
\enq
By imitating the counting of $n^{pq}_j$, we introduce the partition function of
block $p$,
\beqy
Z_{l^p_k}^{p}(\tau)&:=&\frac{1}{pq^2}\left(
G_0^{p}(\tau)+\zeta_{p}^{-(l^p_k)}G_1^{p}(\tau)+\cdots 
+\zeta_{p}^{-(l^p_k)(p-1)}G_{p-1}^p(\tau)
\right)
\no\\
&&
+Z^{pq}_{k}(\tau)
,0\le l^p_k \le p-1, 
\enqy
where $l^p_k$ satisfies $l^p_kq^2\equiv 2qk \mbox{ mod }2pq$.\\
The partition function of block $q$ are similarly introduced.\\
Finally for block $1$, we introduce
\beqy
Z^1_t(\tau)&=&\frac{1}{p^2q^2}G^1_t(\tau)+Z^p_0(\tau)+Z^q_0(\tau)-Z^{pq}_0(\tau)\no\\&=&\frac{1}{p^2q^2}\sum_{0\le a,b,d \in{\bf Z} \atop ad=pq,b<d}
dG(\frac{a\tau+b}{d}).
\label{zpqt}
\enqy
\beqy
Z_{SU(pq)/{\bf Z}_{pq}}(\tau)&:=&\sum_{v_N}Z_{v_N}(\tau)
\no\\&=&
n_0^{pq}Z_0^{pq}(\tau)+\cdots+n_{pq-1}^{p^m}Z_{pq-1}^{p^m}(\tau)+n_0^{p}Z_0^{p}(\tau)+\cdots
+n_{p-1}^{p}Z_{p-1}^{p}(\tau)
\no\\
&&+n_0^{q}Z_0^{q}(\tau)+\cdots
+n_{q-1}^{q}Z_{q-1}^{q-1}(\tau)
+n_t^1Z_t^1(\tau)
\no\\&=&\frac{1}{p^2q^2}\sum_{0\le a,b,d \in{\bf Z} \atop ad=pq,b<d}
d^{12}(\mbox{gcd}(b,d))^{11}G(\frac{a\tau+b}{d}).
\label{zpqd}
\enqy
(\ref{zpqt}) and (\ref{zpqd}) are the same as the result in \cite{lozano}.
\section{Stringy Point of View}
\label{sec:3}
\setcounter{equation}{0}
In this section, we consider the origin of the duality conjecture on the 
${\cal N}=4$ Yang-Mills theory from the stringy point of view. 
Next, we estimate the general structure of the partition function
in the $D, E$ case, based on this observation. In \cite{vafa}, the duality 
conjecture on gauge theory on ${\bf R}^{4}$ is interpreted as the 
$T$-duality of type II A string theory compactified on $T^{2}\times (ALE 
\;\mbox{space})$. In this context, gauge group is determined by the type 
of the $ALE$ space, which is classified by the $ADE$ sub-group of $SU(2)$, 
and the gauge coupling is identified with the deformation parameter $\tau$ 
of the complex structure of $T^{2}$. In this way, $S$-duality transformation 
of the Yang-Mills theory is derived as the $T$-duality transformation of 
$T^{2}$.

Now, we apply this interpretation to the ${\cal N}=4$ $ADE$ Yang-Mills theory  
on $K3$ surface. We can naturally see that the corresponding string theory 
is the type II A string theory compactified on $K3\times T^{2}\times (ALE 
\;\mbox{space})$. In this case, we have to note that both $K3$ and $ALE$ 
space are hyper K\"ahler surfaces. Therefore, we can choose a  
surface of which we take zero-volume limit. If we shrink the $T^{2}\times (ALE 
\;\mbox{space})$, we obtain the ${\cal N}=4$ $ADE$ Yang-Mills theory  
on $K3$ surface. Then what happens if we shrink $T^{2}\times K3$ first ?
We can speculate naturally that the resulting effective theory is the 
${\cal N}=4$ gauge theory on $ALE$ space. Then what we have to do next is 
determination of the gauge group. This can be derived in the following 
way. Let us assume that $K3$ surface has the elliptic fibration. Generically, 
elliptic fibered $K3$ has $24$ singular fibers. Each singular fiber has a 
simple nodal singularity and generate $U(1)$ gauge symmetry. Therefore, 
we can see that the resulting gauge group is $U(1)^{24}$. This result is 
quite natural if we remember the duality between type II A string theory 
on $K{3}$ and heterotic string theory on $T^{4}$ \cite{hj}. 
Since ${\cal N}=4$ gauge 
theory on $K{3}$ does not depend on the complex structure of $K{3}$, 
we don't have to consider the gauge symmetry enhancement. In sum, 
we have obtained the following picture:
\beq
(\mbox{$ADE$ Yang-Mills theory  
on $K3$})\simeq (\mbox{$U(1)^{24}$ Yang-Mills theory  
on $ADE$ $ALE$ space}).
\label{dual} 
\enq 
Now, we derive the partition function of ${\cal N}=4$ $U(1)$ gauge theory 
on the $ALE$ space. We can obtain this theory as a M5-brane world volume theory 
wrapped on $T^{2}\times (ALE\;\mbox{space})$. Appearance of $U(1)$ gauge 
symmetry corresponds to the case of singly wrapped brane. According to 
\cite{m-v},\cite{bonelli}, the partition function of singly wrapped M5-brane 
is determined by the $H^{2}$ lattice and Euler number of the $ALE$ space. 
Let us denote the $ALE$ space associated with $ADE$ group ${\cal G}$ 
by $ALE({\cal G})$. It is well known that 
\begin{equation}
H^{2}(ALE({\cal G}),{\bf Z})=-(\Lambda({\cal G})),\;\;
\chi(ALE({\cal G}))=r+1,
\end{equation}
where $-(\Lambda({\cal G}))$ represents the root lattice whose metric is given 
by the minus of the standard one. 
Then the partition function $Z^{ALE({\cal G})}_{U(1)}(\tau)$ is given by 
\beqy
Z^{ALE({\cal G})}_{U(1)}(\tau)&=&\frac{\theta_{\cal G}^{0}(\tau)}
{\eta(\tau)^{r+1}},\no\\
\theta_{\cal G}^{0}(\tau)&:=&\sum_{m\in {\bf Z}^{r}}
q^{\frac{1}{2}{}^t(m)C_{\cal G}(m)}.
\label{dual2}
\enqy
where $q:=\exp(2\pi i \tau)$, and $C_{\cal G}$ is Cartan Matrix of  
${\cal G }$. Notice that the above result is compatible with the result of 
Nakajima \cite{naka}, that asserts that direct sum of the cohomology of 
the moduli space of $U(1)$ instantons on $ALE({\cal G})$ is decomposed 
into level $1$ representations of the affine Lie algebra of ${\cal G}$.
Combining (\ref{dual}) and (\ref{dual2}), we estimate (at least as the first 
approximation) that the partition 
function $Z^{K3}_{\cal G}(\tau)$ of ${\cal N}=4$ ${\cal G}$ Yang-Mills theory  
on $K3$ surface has the following form:
\begin{equation}
Z^{K3}_{\cal G}(\tau)=\biggl(\frac{\theta_{\cal G}^{0}(\tau)}
{\eta(\tau)^{r+1}}\biggr)^{24}.
\label{1st}
\end{equation}  
\section{Denominator Identity and Partition Function for $D, E$ Gauge 
Groups on $K3$}
\label{sec:4}
\setcounter{equation}{0}
In this section, we turn into another suggestive result useful for 
determination of the partition function of 
$D,E$ gauge theory.
In our previous work \cite{jin2}, we came across a mysterious identity:
\beq
\frac{1}{\eta(\frac{\tau}{N})}=\frac{\theta^2_{A_{N-1}}(\tau)}{\eta(\tau)^N},
\label{deno}
\enq
where $\theta^2_{A_{N-1}}(\tau)$ is a kind of $A_{N-1}$ theta function
which is explicitly defined for general $ADE$ groups as follows: 
 \begin{defi}
For rank $r$ gauge group ${\cal G}$, which has dual Coxeter number $h$,
half of the sum of the positive roots $\rho$ 
and roots $\{ \alpha_k\}$
(see \cite{kac}),
we define the following function  
\begin{eqnarray}
\theta_{\cal G}^2(\tau)&:=& 
\sum_{m\in {\bf Z}^r}q^{\frac{1}{2}{}^t(m+\frac{\rho}{h})C_{\cal G}(m+\frac{\rho}{h})},
\end{eqnarray}
where $q:=\exp(2\pi i \tau)$.
\end{defi}
Since the identity (\ref{deno}) originates from the celebrated denominator 
identity of the affine Lie algebra, it is straightforward  
to generalize (\ref{deno}) to $D,E$ case:
\begin{thm}(Macdonald, Kac) \cite{mac}, \cite{kac}
\beq
\sum_{m\in {\bf Z}^r}q^{\frac{1}{2}{}^t(m+\frac{\rho}{h})C_{\cal G}(m+\frac{\rho}{h})}
=q^{\frac{1}{24}\frac{\dim({\cal G})}{h}}
\prod_{n=1}^\infty (1-q^{\frac{n}{h}})^r\prod_{\alpha_k}(1-q^{\frac{n}{h}}
\zeta_h^{<\rho,\alpha_k>}),
\label{deno2}
\enq
where $\zeta_h:=\exp(2\pi i /h)$.
\end{thm} 
Furthermore as will be shown in appendix B, the r.h.s. of (\ref{deno2}) 
can be expressed as $\eta$-product. 

With the consideration in the previous section, we had better regard this 
identity as the one that interpolate the previous estimation of the 
partition function in (\ref{1st}) to the well-known form given by 
$\eta$-product. Of course, we have to note that the theta function appearing 
in (\ref{deno2}) is not $\theta^{0}_{\cal G}(\tau)$ but 
$\theta^{2}_{\cal G}(\tau)$. This 
subtle translation by $\frac{\rho}{h}$ in the theta sum is puzzling, and  
seems to be caused by boundary condition of the $ALE$ space. This point 
should be pursued further, but instead, we consider here how to use the 
identity (\ref{deno2}). To this aim, we look back at the construction of 
$A_{N}$ partition function given in Section 2. In Section 2, we used 
$G_{0}(\tau)=({\theta^2_{A_{N-1}}(\tau)}/{\eta(\tau)^N})^{24}$
as the primary building block and summed up the $SL(2,{\bf Z})$ orbit of 
$({\theta^2_{A_{N-1}}(\tau)}/{\eta(\tau)^N})^{24}$ so 
that the whole partition functions satisfy the duality conjecture.
Therefore, we expect that the identity (\ref{deno2}) tells us how to 
modify the first approximation (\ref{1st}) of the $D,\; E$ partition 
function on $K3$.  The reason why the $A_{N}$ partition function is written 
in the form of $\eta$-product comes from the fact that the moduli space 
of $SU(N)$ instanton on $K3$ is described by Hilbert scheme of points on 
$K3$. This fact can be derived by Fourier-Mukai transform of semi-stable 
sheaves on $K3$. Since we treat $D, E$ type instanton on the same $K3$, we 
speculate that some analogous structure may exist in the moduli space of 
$D, E$ type instanton on $K3$. In other words, the primary building block 
of the $D, E$ partition function should take the form of $\eta$-product.
Here, we show the table of the $\eta$-products obtained from the identity 
(\ref{deno2}).
\begin{table}[h]
\begin{center}
\begin{tabular}{|c|c|c|}
\hline
${\cal G}$& $\theta^2_{\cal G}(\tau)/\eta(\tau)^{r+1}$\\
\hline
$A_{r}$ & {\Large$\frac{1}{\eta(\frac{\tau}{r+1})}$} \\
\hline
$D_{r}$ &  {\Large$\frac{\eta(\frac{\tau}{r-1})}{\eta(\frac{\tau}{2})\eta(\frac{\tau}{2r-2})}$}\\
\hline
$E_6$ &  {\Large$\frac{\eta(\frac{\tau}{4})\eta(\frac{\tau}{6})}{\eta(\frac{\tau}{2})\eta(\frac{\tau}{3})\eta(\frac{\tau}{12})}$}\\
\hline
$E_7$ &  {\Large$\frac{\eta(\frac{\tau}{6})\eta(\frac{\tau}{9})}{\eta(\frac{\tau}{2})\eta(\frac{\tau}{3})\eta(\frac{\tau}{18})}$}\\
\hline
$E_8$ &  {\Large$\frac{\eta(\frac{\tau}{6})\eta(\frac{\tau}{10})\eta(\frac{\tau}{15})}{\eta(\frac{\tau}{2})\eta(\frac{\tau}{3})\eta(\frac{\tau}{5})\eta(\frac{\tau}{30})}$}\\
\hline
\end{tabular}
\end{center}
\end{table}
With these consideration, we choose for 
the primary function of $D,E$ gauge groups,
\beq
G_0(\tau)
=\left(\frac{\theta^2_{\cal G}(\tau)}{\eta(\tau)^{r+1}}\right)^{24}.
\enq
Because these functions
have the form of $\eta$-product,
we can repeat the same process in Sec.2.4 for these primary functions.
As a result, we obtain
the partition function for $D,E$ gauge groups on $K3$,
which satisfy Montonen-Olive duality (\ref{m-o}).
Though our computation stems from a wild speculation, it is quite 
non-trivial that the resulting partition functions satisfy Montonen-Olive 
duality condition in all the cases we test.   

Before moving to the individual case, we will sketch the processes of 
the determination of $D,E$ partition function on $K3$.
First we have to prepare the sets of $G$ functions.
Here we point out that each $D,E$ theory has
the corresponding $SU(N)$ theory characterized by its center.
We prepare $G$ functions so that all the $G$ functions reproduce
the corresponding $SU(N)$ structure.
Most of $G$'s are obtained by $SL(2,{\bf Z})$ transformation of $G_0(\tau)$,
but some $G$ functions are added in order to satisfy
the above condition. 
Basically one can obtain $D,E$ partition function on $K3$
by imposing Montonen-Olive duality on some combination of $G$ functions.
However we also introduce the intermediate functions ${\tilde G}$,
which are some combination of $G$ functions.
Here $SL(2,{\bf Z})$ transformation of ${\tilde G}$ functions is 
the same as 
that of $G$ functions of
the corresponding $SU(N)$ theory.

\subsection{$D_{2N} $}
$D_{2N}$ is dual to $D_{2N}/{\bf Z}_2\times {\bf Z}_2$.
Thus we introduce 't Hooft flux $v\in H^2(K3, {\bf Z}_2\times {\bf Z}_2)$.
As we have already saw in Sec.2.3, $Z_v(\tau)$ are classified 
by degenerated ${\bf Z}_2$. In this subsection, we try to derive 
$D_2$ and $D_4$ explicitly.

{\large $D_2$}\\
Following the $SU(2)$ case, we first introduce the $G$ functions
obtained from transforming $G_0(\tau)$ by $SL(2,{\bf Z})$.
\beq
G_j(\tau):=\left(
\frac{\eta(\tau)}{\eta(\frac{\tau+j}{2})^2}
\right)^{24},j=0,1,
\enq
\beq
H_0(\tau):=\left(
\frac{\eta(\tau)}{\eta(2\tau)^2}
\right)^{24}.
\enq
To define the partition function for even type ($v^2\equiv 0$) and odd type($v^2\equiv 2 $), we introduce 
\beq
Z_j(\tau):=\frac{1}{4}\left(
G_0(\tau)+(-1)^jG_1(\tau)
\right),j=0,1,\label{zd2j}
\enq
We also introduce the partition function of trivial type,
\beq
Z_t(\tau):=\frac{1}{16}H_0(\tau)+Z_0(\tau).\label{zd2t}
\enq
By using (\ref{zd2j}) and (\ref{zd2t}) and the numbers $\{ n_j\} $ in Sec.2.3,
we obtain
\beq
Z_{D_2/{\bf Z}_2\times {\bf Z}_2}(\tau):=n_0Z_0(\tau)+n_1Z_1(\tau)+n_tZ_t(\tau).\label{zd2d}
\enq
(\ref{zd2t}) and (\ref{zd2d}) surely satisfy Montonen-Olive duality (\ref{m-o}),
\beq
Z_t\left(-\frac{1}{\tau}\right)=4^{-11}\left(\frac{\tau}{i}\right)^{-12}
Z_{D_2/{\bf Z}_2\times {\bf Z}_2}(\tau).\label{zd2dd}
\enq
Note that the coefficients of $G$'s in (\ref{zd2j}) and (\ref{zd2t}) are 
determined by imposing (\ref{zd2dd}).

{\large $D_4$}\\
Following the $D_2$ case, we first introduce the $G$ functions
obtained from transforming $G_0(\tau)$ by $SL(2,{\bf Z})$.
\beq
G_j(\tau):=\left(
\frac{\eta(\frac{\tau+j}{3})}{\eta(\frac{\tau+j}{2})\eta(\frac{\tau+j}{6})}
\right)^{24},j=0,\ldots,5,
\enq
\beq
H_j(\tau):=\left(
\frac{\eta(\frac{\tau+j}{3})}{\eta(2\tau)\eta(\frac{2\tau+2j}{3})}
\right)^{24},j=0,\ldots,2,
\enq
\beq
I_j(\tau):=\left(
\frac{\eta(3\tau)}{\eta(\frac{\tau+j}{2})\eta(\frac{3\tau+3j}{2})}
\right)^{24},j=0,1,
\enq
\beq
J_0(\tau):=\left(
\frac{\eta(3\tau)}{\eta(2\tau)\eta(6\tau)}
\right)^{24}.
\enq
Contrary to the $D_2$ case, we will introduce 
the intermediate
${\tilde G }(\tau)$ functions,
so that we pick up the powers $q^{(1/2){\bf Z}}$
from the $G$ functions having $q^{(1/6){\bf Z}}$ powers,
\beq
{\tilde G}_j(\tau):=G_j(\tau)+G_{2+j}(\tau)+G_{4+j}(\tau)+I_j(\tau),j=0,1,
\enq
\beq
{\tilde H}_0(\tau):=H_0(\tau)+H_1(\tau)+H_2(\tau)+J_0(\tau).
\enq
Once we introduce ${\tilde G}(\tau)$'s, we only follow the same processes as $D_2$ case. Actually the modular property of ${\tilde G}_j(\tau)$ and ${\tilde H}_0(\tau)$ is completely the same as that of $G_j(\tau)$ and $H_0(\tau)$ in $D_2$. 
\beq
Z_j(\tau):=\frac{1}{4}\left(
{\tilde G}_0(\tau)+(-1)^j{\tilde G}_1(\tau)
\right),j=0,1,
\enq
\beq
Z_t(\tau):=\frac{1}{16}{\tilde H}_0(\tau)+Z_0(\tau),\label{zd4t}
\enq
\beq
Z_{D_4/{\bf Z}_2\times {\bf Z}_2}(\tau):=n_0Z_0(\tau)+n_1Z_1(\tau)+n_tZ_t(\tau),\label{zd4d}
\enq
(\ref{zd4t}) and (\ref{zd4d}) also satisfy Montonen-Olive duality (\ref{m-o}),
\beq
Z_t\left(-\frac{1}{\tau}\right)=4^{-11}\left(\frac{\tau}{i}\right)^{-12}
Z_{D_4/{\bf Z}_2\times {\bf Z}_2}(\tau).
\enq
\subsection{$D_{2N+1} $}
$D_{2N+1}$ is dual to $D_{2N+1}/{\bf Z}_4$. 
Thus we introduce 't Hooft flux $v\in H^2(K3, {\bf Z}_4)$.
As we have already saw in Sec.2.3, $Z_v(\tau)$ are classified by ${\bf Z}_4$.
This structure is the same as $A_3$ case. Indeed
\[
D_3=A_3.
\]
In this subsection, we try to derive the  
$D_5$ partition function explicitly.

{\large $D_5$}\\
Following the $SU(4)$ case, we first introduce the $G$ functions
obtained from transforming $G_0(\tau)$ by $SL(2,{\bf Z})$,
and add some $G$ functions so that all $G$ functions form 
the modular property of $SU(4)$.
\beq
G_j(\tau):=\left(
\frac{\eta(\frac{\tau+j}{4})}{\eta(\frac{\tau+j}{2})\eta(\frac{\tau+j}{8})}
\right)^{24},j=0,\ldots,7,
\enq
\beq
H_j(\tau):=\left(
\frac{\eta(\frac{2\tau+j}{2})}{\eta(2\tau)\eta(\frac{2\tau+j}{4})}
\right)^{24},j=0,\ldots,3,
\enq
\beq
I_0(\tau):=\left(
\frac{\eta(4\tau)}{\eta(2\tau)\eta(\frac{4\tau+1}{2})}
\right)^{24},
\enq
\beq
J_0(\tau):=\left(
\frac{\eta(4\tau)}{\eta(2\tau)\eta(8\tau)}
\right)^{24},
\enq
\beq
K_j(\tau):=\left(
\frac{\eta(2\tau)}{\eta(\frac{2\tau+j}{2})^2}
\right)^{24},j=0,1,
\enq
\beq
L_j(\tau):=\left(
\frac{\eta(\frac{\tau+j}{2})}{\eta(\frac{\tau+j}{4})^2}
\right)^{24},j=0,\ldots,3,
\enq
\beq
M_0(\tau):=\left(
\frac{\eta(2\tau)}{\eta(4\tau)^2}
\right)^{24}.
\enq
Precisely speaking, $G_j,(j=0,\ldots,7),H_1,H_3,I_0,J_0$ are obtained from 
transforming $G_0(\tau)$ by $SL(2,{\bf Z})$, and 
$H_0,H_2,K_j,(j=0,1),L_j,(j=0,\ldots,3),M_0$
are obtained from requiring the modular property of $SU(4)$.\\
For the same reason as  the $D_4$ case, we will introduce 
the intermediate
${\tilde G }(\tau)$ functions,
so that we pick up the powers $q^{(1/4){\bf Z}}$ 
from $G$ functions, 
\beq
{\tilde G}_j^4(\tau):=G_j(\tau)+G_{4+j}(\tau)+2^{12}L_j(\tau),j=0,\ldots,3\label{g4t}
\enq
\beq
{\tilde G}_j^2(\tau):=H_j(\tau)+H_{2+j}(\tau)+2^{12}K_j,j=0,1,\label{g2t}
\enq
\beq
{\tilde G}_t^1(\tau):=2^{12}I_0(\tau)+J_0(\tau)+M_0(\tau).\label{gtt}
\enq
The modular property of ${\tilde G}_j^4(\tau)$, 
${\tilde G}_j^2(\tau)$ and ${\tilde G}_t^1(\tau)$ is completely 
the same as that of $G_j^4(\tau)$, $G_j^2(\tau)$and $G_t^1(\tau)$ in $SU(4)$
respectively.
Mysterious $2^{12}$ factor in (\ref{g4t}), (\ref{g2t}) and (\ref{gtt}) is
needed for this equivalence.
By using ${\tilde G}(\tau)$'s, we follow the same processes as 
the $SU(4)$ case.
\beq
Z_j^{4}(\tau):=\frac{1}{4}\left(
{\tilde G}_0^4(\tau)+\zeta_4^{-j}{\tilde G}_1^4(\tau)+\cdots +\zeta_4^{-j3}{\tilde G}_{3}^1(\tau)
\right),j=0,\ldots,3,
\enq
\beq
Z_j^{2}(\tau):=
\frac{1}{8}\left(
{\tilde G}_0^2(\tau)+(-1)^j{\tilde G}_{1}^2(\tau)
\right)+Z^{4}_0(\tau)
,j=0,1,
\enq
\beq
Z_t^1(\tau):=\frac{1}{16}{\tilde G}_t^1(\tau)+Z_0^{2}(\tau),\label{zd5t}
\enq
\beq
Z_{D_5/{\bf Z}_4}(\tau):=n_0^{4}Z_0^{4}(\tau)+\cdots+n_{3}^{4}Z_{N-1}^{4}(\tau)+n_0^{2}Z_0^{2}(\tau)+n_1^{2}Z_1^{2}(\tau)+n_t^1Z_t^1(\tau).\label{zd5d}
\enq
(\ref{zd5t}) and (\ref{zd5d}) indeed  
satisfy Montonen-Olive duality (\ref{m-o})
in the same way as $SU(4)$ case,
\beq
Z_t^1\left(-\frac{1}{\tau}\right)=4^{-11}\left(\frac{\tau}{i}\right)^{-12}
Z_{D_5/{\bf Z}_4}(\tau).
\enq

\subsection{$E_{6,7,8}$}
{\large $E_6$}\\
$E_6$ is dual to $E_6/{\bf Z}_3$. 
Thus we introduce 't Hooft flux $v\in H^2(K3, {\bf Z}_3)$.
As we already saw in Sec.2.3, $Z_v(\tau)$ are classified by ${\bf Z}_3$.
Explicit $G$ functions are written in Appendix C.
Following the $D_4$ and $D_5$ case, we introduce the intermediate functions,
\beq
{\tilde G}_j(\tau):=\sum_{k=0}^3G_{3k+j}(\tau)
+\sum_{k=0}^1H_{3k+3-j}(\tau)+\sum_{k=0}^1N_{3k+j}(\tau)+J_{3-j}(\tau),j=0,1,2,
\enq
\beq
{\tilde H}_0(\tau):=\sum_{k=0}^3I_k(\tau)+\sum_{k=0}^1K_k(\tau)
+\sum_{k=0}^1M_k(\tau)+L_0(\tau).
\enq
By using ${\tilde G}(\tau)$'s, we follow the same process as the  $SU(3)$ case.
\beq
Z_j(\tau):=\frac{1}{3}\left(
{\tilde G}_0(\tau)+\zeta_3^{-j}{\tilde G}_1(\tau)+\zeta_3^{-j2}{\tilde G}_{2}(\tau)
\right),j=0,1,2,
\enq
\beq
Z_t(\tau):=\frac{1}{9}{\tilde H}_0(\tau)+Z_0(\tau),\label{ze6t}
\enq
\beq
Z_{E_6/{\bf Z}_3}(\tau):=n_0Z_0(\tau)+n_{1}Z_{1}(\tau)+n_{2}Z_{2}(\tau)+n_tZ_t(\tau).\label{ze6d}
\enq
(\ref{ze6t}) and (\ref{ze6d}) surely satisfy Montonen-Olive duality (\ref{m-o})
in the same way as the $SU(3) $ case,
\beq
Z_t\left(-\frac{1}{\tau}\right)=3^{-11}\left(\frac{\tau}{i}\right)^{-12}
Z_{E_6/{\bf Z}_3}(\tau).
\enq
{\large $E_7$}\\
By repeating the similar processes to $E_6$ case, the results of $E_7$ 
are given by
\begin{eqnarray}
{\tilde G}_j(\tau)&:=&\sum_{k=0}^8G_{2k+j}(\tau)+\sum_{k=0}^2I_{2k+j}(\tau)
+\sum_{k=0}^2N_{2k+j}(\tau)+K_j(\tau)
,j=0,1,
\end{eqnarray}
\begin{eqnarray}
{\tilde H}_0(\tau)&:=&\sum_{k=0}^8H_k(\tau)+\sum_{k=0}^3J_k(\tau)
+\sum_{k=0}^2M_k(\tau)+L_0(\tau).
\end{eqnarray}
Corresponding $SU(N)$ theory is $SU(2)$. Thus we define
\beq
Z_j(\tau):=\frac{1}{2}\left(
{\tilde G}_0(\tau)+(-1)^{-j}{\tilde G}_1(\tau)
\right),j=0,1,
\enq
\beq
Z_t(\tau):=\frac{1}{4}{\tilde H}_0(\tau)+Z_0(\tau),\label{ze7t}
\enq
\beq
Z_{E_7/{\bf Z}_2}(\tau):=n_0Z_0(\tau)+n_{1}Z_{1}(\tau)+n_tZ_t(\tau).\label{ze7d}
\enq
(\ref{ze7t}) and (\ref{ze7d}) satisfy Montonen-Olive duality (\ref{m-o})
in the same way as the $SU(2)$ case,
\beq
Z_t\left(-\frac{1}{\tau}\right)=2^{-11}\left(\frac{\tau}{i}\right)^{-12}
Z_{E_7/{\bf Z}_2}(\tau).
\enq
{\large $E_8$}\\
Since $E_8$ is self-dual group, $Z_t(\tau)$ turns out to be a modular 
form of weight $-12$. 
In this case, we introduce $G$ functions obtained from transforming $G_0$ by 
$SL(2,{\bf Z})$.

\beq
G_j(\tau):=\left(
\frac{\eta(\frac{\tau+j}{6})\eta(\frac{\tau+j}{10})\eta(\frac{\tau+j}{15})}{\eta(\frac{\tau+j}{2})\eta(\frac{\tau+j}{3})\eta(\frac{\tau+j}{5})\eta(\frac{\tau+j}{30})}
\right)^{24},j=0,\ldots,29,
\enq
\beq
H_j(\tau):=\left(
\frac{\eta(\frac{2\tau+2j}{3})\eta(\frac{2\tau+2j}{5})\eta(\frac{\tau+j}{15})}{\eta(2\tau)\eta(\frac{\tau+j}{3})\eta(\frac{\tau+j}{5})\eta(\frac{2\tau+2j}{15})}
\right)^{24},j=0,\ldots,14,
\enq
\beq
I_j(\tau):=\left(
\frac{\eta(\frac{3\tau+3j}{2})\eta(\frac{\tau+j}{10})\eta(\frac{3\tau+3j}{5})}{\eta(\frac{\tau+j}{2})\eta(3\tau)\eta(\frac{\tau+j}{5})\eta(\frac{3\tau+3j}{10})}
\right)^{24},j=0,\ldots,9,
\enq
\beq
J_j(\tau):=\left(
\frac{\eta(\frac{\tau+j}{6})\eta(\frac{5\tau+5j}{2})\eta(\frac{5\tau+5j}{3})}{\eta(\frac{\tau+j}{2})\eta(\frac{\tau+j}{3})\eta(5\tau)\eta(\frac{5\tau+5j}{6})}
\right)^{24},j=0,\ldots,5,
\enq
\beq
K_j(\tau):=\left(
\frac{\eta(6\tau)\eta(\frac{2\tau+2j}{5})\eta(\frac{3\tau+3j}{5})}{\eta(2\tau)\eta(3\tau)\eta(\frac{\tau+j}{5})\eta(\frac{6\tau+6j}{5})}
\right)^{24},j=0,\ldots,4,
\enq
\beq
L_j(\tau):=\left(
\frac{\eta(\frac{2\tau+2j}{3})\eta(10\tau)\eta(\frac{5\tau+5j}{3})}{\eta(2\tau)\eta(\frac{\tau+j}{3})\eta(5\tau)\eta(\frac{10\tau+10j}{3})}
\right)^{24},j=0,1,2,
\enq
\beq
M_j(\tau):=\left(
\frac{\eta(\frac{3\tau+j}{2})\eta(\frac{5\tau+j}{2})\eta(15\tau)}{\eta(\frac{\tau+j}{2})\eta(3\tau)\eta(5\tau)\eta(\frac{15\tau+j}{2})}
\right)^{24},j=0,1,
\enq
\beq
N_0(\tau):=\left(
\frac{\eta(6\tau)\eta(10\tau)\eta(15\tau)}{\eta(2\tau)\eta(3\tau)\eta(5\tau)\eta(30\tau)}
\right)^{24},
\enq
Each $G$ function is transformed to each other by $SL(2,{\bf Z})$
with no extra factor except for $(\frac{\tau}{i})^{-12} $.
Therefore, there is no need to introduce the intermediate functions.
The partition function is simply given by adding all the $G$ functions. 
\beqy
Z_t(\tau)&:=&\sum_{j=0}^{29}G_j(\tau)+\sum_{j=0}^{14}H_j(\tau)+\sum_{j=0}^9I_j(\tau)+\sum_{j=0}^5J_j(\tau)+\sum_{j=0}^4K_j(\tau)+\sum_{j=0}^2L_j(\tau)
\no\\&&
+\sum_{j=0}^1M_j(\tau)
+N_0(\tau).
\no\\&&\label{ze8t}
\enqy
Therefore, (\ref{ze8t}) is surely self-dual or a modular form,
\beq
Z_t\left(-\frac{1}{\tau}\right)=\left(\frac{\tau}{i}\right)^{-12}
Z_{t}(\tau).
\enq
Indeed $Z_t(\tau)$ can be expressed by the well-known modular forms 
as follows:
\beqy
Z_t(\tau)&=&\frac{1}{\eta(\tau)^{24}}\left(j^8(\tau)-5976j^7(\tau)+14049204j^6(\tau)-16450492296j^5(\tau)\right.
\no\\&&
+10006744823442j^4(\tau)
-2995100782701144j^3(\tau)
+373127947258066100j^2(\tau)
\no\\&&\left.
-12808385327808647208j(\tau)+26763599994092029512\right),
\enqy
where $j(\tau)$
is the famous modular $j$-function \cite{miyake}.
Notice that
\beq 
j(\tau)=\chi_{1,0}(\tau)^3
\enq
where $\chi_{1,0}(\tau)$ is level one $E_8$ character.
In this way, $Z_t(\tau)$ can be expressed only in terms of $\chi_{1,0}(\tau)$
and $\eta(\tau)$.
This fact ensures that $Z_t(\tau)$ is a good candidate for 
the $E_8$ partition function on $K3$.

\section{Holomorphic Anomaly ?}
\label{sec:6}
\setcounter{equation}{0}

In this section, we discuss that our partition function for $D,E$
gauge group is not sufficient from the point of view of the gap condition,
and search for the improvement analogous to the holomorphic anomaly 
on $\frac{1}{2}K3 $ \cite{m-v}.

First, we derive the gap condition for $ADE$ gauge groups by using 
the famous Atiyah-Hitchin-Singer dimension formula for
the moduli space of irreducible anti-self-dual connections 
with $ADE$ gauge group ${\cal G}$ 
on $X$ \cite{AHS}.
\beq
\dim{\cal M}^{\cal G}_k=4h({\cal G})k-\dim{\cal G}\frac{\chi(X)+\sigma(X)}{2},
\enq 
where $h({\cal G})$ is the dual Coxeter number, $\dim{\cal G}$ is the dimension
of ${\cal G}$ and $k$ is the instanton number. $\chi(X)$ and $\sigma(X)$
are Euler number and signature of $X$ respectively.
In our case of $X=K3$, this formula reduces to 
\beq
\dim{\cal M}^{\cal G}_k=4h({\cal G})k-4\dim{\cal G}.
\enq 
Here we point out that $h({\cal G}),\dim({\cal G})$ and $r$
are related by the following formula:
\beq
\dim({\cal G})=rh({\cal G})+r,h({\cal G})>r.
\enq
Therefore,  
\beq
\dim{\cal M}^{\cal G}_k=4h({\cal G})(k-r)-4r.
\enq 
$ \dim{\cal M}^{\cal G}_k\ge 0$ means $k\ge r+1$.
That is, the moduli space of irreducible ASD connections with 
$ADE$ gauge group on $K3$ can exist only for $k\ge r+1$.
This condition restricts the form of the partition function. 
Remember that the partition function has the form:
 \begin{equation}
Z^X_v(\tau):= q^{-{\frac{(r+1)\chi(X)}{24}}}\sum_k \chi({\cal M}(v,k))q^k
\;\;\;.\label{zxvtd}
\end{equation}
In our case, we consider the case of  $X=K3$ and of trivial flux, 
which corresponds to
$ {\cal M}(0,k)={\cal M}^{\cal G}_k$. Thus we obtain
\begin{equation}
Z_t(\tau) = q^{-(r+1)}\sum_k \chi({\cal M}^{\cal G}_k)q^k
\;\;\;.
\label{zxvtdd}
\end{equation}
This form and the condition $k\ge r+1 $ requires (\ref{zxvtdd})
to start from ${\cal O}(q^0) $ term.
Furthermore, it is well-known that the trivial connection always exists
\cite{vafa-witten}.
This means that the contribution from $k=0$ appears in (\ref{zxvtdd}) for 
any gauge group.
Therefore, 
forbidden $q$ powers in (\ref{zxvtdd}) are $q^{-r},\ldots,q^{-1} $,
that is $1\le k \le r$.
This is the gap condition.

The $SU(N)$ partition function on $K3$
surely satisfies this gap condition \cite{vafa-witten}.
However our $D,E$ partition function on $K3$
does not satisfy this gap condition. 
To discuss this problem, we will pick up the $E_8$
partition function on $K3$.
The other $D,E$ partition functions on $K3$ have the same 
problem as the $E_8$ case.
First we expand $Z_t(\tau)$ of $E_8$:
\beqy
Z_t(\tau)&=&q^{-9}+24q^{-7}+24q^{-6}+324q^{-5}+600q^{-4}+3500q^{-3}+10008q^{-2}
+22890627q^{-1}
\no\\
&&
+134931884553792+1697352144366449440512q
\no\\&&
+177293857723945037797591920q^2
+2444259891832216559852811815364q^3
\no\\&&
+10159099239429369126457917291280224q^4
+\cdots.
\enqy \
What is the origin of $q^{-7},q^{-6},\ldots,q^{-1} $ ?
We can suggest the following two possibilities.
The one possibility is the failure of 
the vanishing theorem \cite{vafa-witten}.
The vanishing theorem ensures the equality (\ref{zxvtdd}).
From this view point, we may have derived the $D,E$ partition function on 
$K3$, which corresponds to the generating function of 
the Euler number of the generalized moduli space 
including the field $A_\mu,B_{\mu\nu}^+$ and $C$ \cite{vafa-witten,laba}.
The second possibility is the appearance of the reducible
connection. 
Atiyah-Hitchin-Singer dimension formula is the formula for
irreducible ASD connection.
Thus if there are reducible connections,
it is no wonder why the negative powers appear in
the partition function.
Actually the above two possibilities are mutually related.
We will discuss the second possibility more in detail here.

First we point out the well-known fact that
there is a holomorphic anomaly in the $SU(N)$
partition function on the manifold $b_2^+\le 1$.
This holomorphic anomaly comes from reducible connections.
We think that our situation is very similar to
the holomorphic anomaly appearing in the $U(N)$ 
partition function on $\frac{1}{2}K3 $
\cite{m-v,E-S}.
For example, let us  
expand the holomorphic part of the $U(2)$ partition function 
$Z_2^{hol.}(\tau)$, which does not include $E_2(\tau)$ terms \cite{m-v,E-S}.
\beq
Z_2^{hol.}(\tau)=\frac{1}{12}q^{-1}-20-11787q-\frac{2132320}{3}q^2+\cdots.
\enq
If one adds $E_2(\tau)$ terms, which are determined by
the holomorphic anomaly equation \cite{m-v},
one can obtain $Z_2(\tau) $, which includes
no ${\cal O}(q^0)$ term.
Looking back at our problem, we may obtain 
the complete partition function of $D,E$ gauge theory 
by adding $E_2(\tau)$ dependent terms, which come from reducible connections. 
Since $E_{2}(\tau)$ behave anomalously under $SL(2,{\bf Z})$ transformation,
each coefficient function should be a modular form of appropriate weight. 
More explicitly, expected full partition function $Z_{t}^{full}(\tau)$ 
should take the following 
form:
\beqy
Z_{t}^{full}(\tau)&=&\frac{1}{\eta(\tau)^{216}}\cdot
\biggl(\sum_{j}(E_{2}(\tau))^{j}\cdot f_{j}(E_{4},E_{6})\biggr),\no\\
Z_{t}(\tau)&=&\frac{f_{0}(E_{4},E_{6})}{\eta(\tau)^{216}},
\enqy  
where $f_{j}(E_{4},E_{6})$ is a modular form of weight $96-2j$.
Of course, the above ansatz has enough degrees of freedom to cancel the 
coefficient of $q^{-7},q^{-6},\ldots,q^{-1}$ of $Z_{t}(\tau)$. 
But we don't have the principle to select the ``right'' $f_{j}(E_{4},E_{6})$
up to now. We expect to find the analogue of holomorphic anomaly equation 
used in \cite{m-v}. Unfortunately, corresponding Seiberg-Witten curve 
is not known and the problem still needs further consideration.

\section{Conclusion and Discussion}
\label{sec:7}
\setcounter{equation}{0}
 
We have determined the holomorphic part of the $D,E$ partition function
on $K3$, by using the $D,E$ blow-up formula.
This partition function completely satisfies Montonen-Olive duality.
 
Remaining problem is determination of the complete partition function of $D,E$  gauge theory on $K3$, which satisfies the gap condition.
We think that the desired complete $D,E$ partition function on $K3$
includes the holomorphic part and the anti-holomorphic part having 
$E_2(\tau)$ terms.
We also expect that 
the anti-holomorphic part of the partition function of $D,E$ 
gauge theory on $K3$ should be determined by some equations, 
which corresponds to the holomorphic anomaly equation on $\frac{1}{2}K3$.
To this aim, an approach from Seiberg-Witten curve seems to be effective 
since the holomorphic anomaly equation was verified from the analysis 
Seiberg-Witten of ${\cal N}=4$ gauge theory. 
\\
{\bf Acknowledgment}\\
We would like to thank T.Eguchi, K.Fukaya, H.Nakajima, K.Ono and K.Yoshioka
for helpful suggestions and useful discussions.
We owe the major part of Sec.5 to the discussions with them.
\newpage
\appendix
\renewcommand{\theequation}{\alph{section}.\arabic{equation}}
\section{Modular property of $SU(3)/{\bf Z}_3$ theory}
\setcounter{equation}{0}
In this article, we only considered the modular property of (\ref{m-o}).
\[
Z_t(-\frac{1}{\tau})=3^{-11}\tau^{-12}(n_0Z_0(\tau)+n_1Z_1(\tau)+n_2Z_2(\tau)+n_tZ_t(\tau)).
\] 
Of course other modular properties in (\ref{vw}) must be considered.
In this section, we derive other modular properties in (\ref{vw})
by counting the orbits for $A_2$ case. 

\subsection{$Z_0(-\frac{1}{\tau}) $}
 \beq
Z_0(-\frac{1}{\tau})=3^{-11}\tau^{-12}(m_0Z_0(\tau)+m_1Z_1(\tau)+m_2Z_2(\tau)+m_tZ_t(\tau)).
\enq
We determine $\{m_j\}$ by using (\ref{vw}).

\beq
\left\{
\begin{array}{c}
m_0=m_0^0+\zeta_3m_0^1+\zeta_3^2m_0^2\\
m_1=m_1^0+\zeta_3m_1^1+\zeta_3^2m_1^2\\
m_2=m_2^0+\zeta_3m_2^1+\zeta_3^2m_2^2\\
m_t=m_t^0,
\end{array}\right.
\enq
where 
\beq
m_j^k=\mbox{number of}~~u^2\equiv 2j \mbox{ mod }6~~{\rm and}~~ V_1\cdot u\equiv k\mbox{ mod }3 ,u\ne 0,
\enq
\[
m_t^0=1.
\]
Here $V_1^2=0$ and $V_1\ne 0$,and $v=V_1$ corresponds to $v^2\equiv 0$ mod $6$ and
$v\ne 0$.
The results are given by
\beq
\left\{
\begin{array}{c}
m_0^0=3^{20}+2\cdot 3^{10}-1\\
m_0^1=3^{20}\\
m_0^2=3^{20}\\
m_0=2\cdot 3^{10}-1,
\end{array}\right.
\enq
\beq
\left\{
\begin{array}{c}
m_1^0=3^{20}-3^{10}\\
m_1^1=3^{20}\\
m_1^2=3^{20}\\
m_1=-3^{10},
\end{array}\right.
\enq
\beq
\left\{
\begin{array}{c}
m_2^0=3^{20}-3^{10}\\
m_2^1=3^{20}\\
m_2^2=3^{20}\\
m_2=-3^{10}.
\end{array}\right.
\enq

\subsection{$Z_1(-\frac{1}{\tau}) $}
 \beq
Z_1(-\frac{1}{\tau})=3^{-11}\tau^{-12}(m_0Z_0(\tau)+m_1Z_1(\tau)+m_2Z_2(\tau)+m_tZ_t(\tau)).
\enq
We determine $\{m_j\}$ by using (\ref{vw}).

\beq
\left\{
\begin{array}{c}
m_0=m_0^0+\zeta_3m_0^1+\zeta_3^2m_0^2\\
m_1=m_1^0+\zeta_3m_1^1+\zeta_3^2m_1^2\\
m_2=m_2^0+\zeta_3m_2^1+\zeta_3^2m_2^2\\
m_t=m_t^0,
\end{array}\right.
\enq
where 
\beq
m_j^k=\mbox{number of}~~u^2\equiv 2j \mbox{ mod }6 ~~{\rm and}~~ (V_1+V_2)\cdot u\equiv k \mbox{ mod }3,u\ne 0,
\enq
\[
m_t^0=1.
\]
Here $(V_1+V_2)^2=2$,and $v=V_1+V_2$ corresponds to $v^2\equiv 2$ mod $6$.
The results are given by
\beq
\left\{
\begin{array}{c}
m_0^0=3^{20}-1\\
m_0^1=3^{20}+3^{10}\\
m_0^2=3^{20}+3^{10}\\
m_0=-3^{10}-1,
\end{array}\right.
\enq
\beq
\left\{
\begin{array}{c}
m_1^0=3^{20}-3^{10}\\
m_1^1=3^{20}\\
m_1^2=3^{20}\\
m_1=-3^{10},
\end{array}\right.
\enq
\beq
\left\{
\begin{array}{c}
m_2^0=3^{20}+3^{10}\\
m_2^1=3^{20}-3^{10}\\
m_2^2=3^{20}-3^{10}\\
m_2=2\cdot3^{10}.
\end{array}\right.
\enq
\section{$ADE$ Blow-up formula }
\setcounter{equation}{0}
\subsection{$A_r $ case}
It is convenient to write roots ${\alpha_k}$ of $A_r$ algebra
by orthonormal basis $\{ e_j \}$.
We write $\rho$ by these basis.
By using these data, we derive $\theta^2_{A_r}(\tau)$ as below.

\beq
\mbox{Roots} \{\alpha_k \}:~~ 
e_i - e_j, ~~(1\le i\ne j \le r+1).
\enq
\beq
h=r+1,
\enq
\beq
\rho=\frac{1}{2}(re_1+(r-2\cdot 1)e_2+\cdots +(r-2\cdot r)e_{r+1}),
\enq
\beq
\zeta_h=\exp(\frac{2\pi i}{h}),
\enq
\begin{eqnarray}
\theta_{A_r}^2(\tau)&=& q^{\frac{1}{24}\frac{r(r+2)}{h}}
\prod_{n=1}^\infty (1-q^{\frac{n}{h}})^r\prod_{\alpha_k}(1-q^{\frac{n}{h}}\zeta_h^{<\rho,\alpha_k>})
\\
&=&
\frac{\eta(\tau)^{r+1}}{\eta(\frac{\tau}{h})},
\end{eqnarray}
\begin{equation}
G_0(\tau):=\left(\frac{\theta^2_{A_r}(\tau)}{\eta(\tau)^{r+1}}\right)^{24}
=\left(\frac{1}{\eta(\frac{\tau}{h})}\right)^{24}.
\end{equation}

\subsection{$D_r $ case}

\beq
\mbox{Roots}\{\alpha_k \}:~ 
\pm e_i \pm e_j, ~~(1\le i\ne j\le r).
\enq
\beq
h=2r-2,
\enq
\beq
\rho=\frac{1}{2}(he_1+(h-2)e_2+\cdots +2e_{r-1}),
\enq
\beq
\zeta_h=\exp(\frac{2\pi i}{h}),
\enq
\begin{eqnarray}
\theta_{D_r}^2(\tau)&=& q^{\frac{1}{24}\frac{r(2r-1)}{h}}
\prod_{n=1}^\infty (1-q^{\frac{n}{h}})^r\prod_{\alpha_k}(1-q^{\frac{n}{h}}\zeta_h^{<\rho,\alpha_k>})
\\
&=&
\frac{\eta(\tau)^{r+1}\eta(\frac{2\tau}{h})}{\eta(\frac{\tau}{2})\eta(\frac{\tau}{h})},
\end{eqnarray}

\begin{equation}
G_0(\tau):=
\left(\frac{\theta^2_{D_r}(\tau)}{\eta(\tau)^{r+1}}\right)^{24}
=\left(\frac{\eta(\frac{2\tau}{h})}{\eta(\frac{\tau}{2})\eta(\frac{\tau}{h})}\right)^{24}.
\end{equation}
\subsection{$E_6 $ case}

\beq
\mbox{Roots } \{ \alpha_k \}: 
\pm e_i \pm e_j ~~(1\le i\ne j\le 5), 
\enq
\beq
\frac{1}{2}(\pm e_1 \pm e_2 \cdots \pm \sqrt{3}e_6),~~\mbox{even $+$'s}.
\enq
\beq
h=12,
\enq
\beq
\rho=\frac{1}{2}(16e_1+6e_2+4e_3+2e_{4}),
\enq
\beq
\zeta_{12}=\exp(\frac{2\pi i}{12}),
\enq
\begin{eqnarray}
\theta_{E_6}^2(\tau)&=& q^{\frac{1}{24}\frac{78}{12}}
\prod_{n=1}^\infty (1-q^{\frac{n}{h}})^6\prod_{\alpha_k}(1-q^{\frac{n}{h}}\zeta_h^{<\rho,\alpha_k>})
\\
&=&
\frac{\eta(\tau)^{7}\eta(\frac{\tau}{4})\eta(\frac{\tau}{6})}{\eta(\frac{\tau}{2})\eta(\frac{\tau}{3})\eta(\frac{\tau}{12})},
\end{eqnarray}
\begin{equation}
G_0(\tau):=\left(
\frac{\eta(\frac{\tau}{4})\eta(\frac{\tau}{6})}{\eta(\frac{\tau}{2})\eta(\frac{\tau}{3})\eta(\frac{\tau}{12})}
\right)^{24}.
\end{equation}

\subsection{$E_7 $ case}

\beq
\mbox{Roots } \{ \alpha_k \}:
\pm e_i \pm e_j ~~(1\le i\ne j\le 6),~~ \pm \sqrt{2}e_7,
\enq
\beq
\frac{1}{2}(\pm e_1 \pm e_2 \cdots \pm \sqrt{3}e_7),~~\mbox{even $+$'s}.
\enq
\beq
h=18,
\enq
\beq
\rho=\frac{1}{2}(26e_1+8e_2+6e_3+4e_{4}+2e_5+\sqrt{2}e_7),
\enq
\beq
\zeta_{18}=\exp(\frac{2\pi i}{18}),
\enq
\begin{eqnarray}
\theta_{E_7}^2(\tau)&=& q^{\frac{1}{24}\frac{133}{18}}
\prod_{n=1}^\infty (1-q^{\frac{n}{h}})^7\prod_{\alpha_k}(1-q^{\frac{n}{h}}\zeta_h^{<\rho,\alpha_k>})
\\
&=&
\frac{\eta(\tau)^{8}\eta(\frac{\tau}{6})\eta(\frac{\tau}{9})}{\eta(\frac{\tau}{2})\eta(\frac{\tau}{3})\eta(\frac{\tau}{18})},
\end{eqnarray}
\begin{equation}
G_0(\tau):=\left(
\frac{\eta(\frac{\tau}{6})\eta(\frac{\tau}{9})}{\eta(\frac{\tau}{2})\eta(\frac{\tau}{3})\eta(\frac{\tau}{18})}
\right)^{24}.
\end{equation}

\subsection{$E_8 $ case}

\beq \mbox{Roots} \{\alpha_k \} :~
\pm e_i \pm e_j ~~(1\le i\ne j\le 8), 
\enq
\beq
\frac{1}{2}(\pm e_1 \pm e_2 \cdots \pm e_8),~~\mbox{even $+$'s}.
\enq
\beq
h=30,
\enq
\beq
\rho=\frac{1}{2}(46e_1+12e_2+10e_3\cdots +2e_{7}),
\enq
\beq
\zeta_{30}=\exp(\frac{2\pi i}{30}),
\enq
\begin{eqnarray}
\theta_{E_8}^2(\tau)&=& q^{\frac{1}{24}\frac{248}{30}}
\prod_{n=1}^\infty (1-q^{\frac{n}{h}})^8\prod_{\alpha_k}(1-q^{\frac{n}{h}}\zeta_h^{<\rho,\alpha_k>})
\\
&=&
\frac{\eta(\tau)^{9}\eta(\frac{\tau}{6})\eta(\frac{\tau}{10})\eta(\frac{\tau}{15})}{\eta(\frac{\tau}{2})\eta(\frac{\tau}{3})\eta(\frac{\tau}{5})\eta(\frac{\tau}{30})},
\end{eqnarray}
\begin{equation}
G_0(\tau):=
\left(\frac{\theta^2_{E_8}(\tau)}{\eta(\tau)^{9}}\right)^{24}
=\left(
\frac{\eta(\frac{\tau}{6})\eta(\frac{\tau}{10})\eta(\frac{\tau}{15})}{\eta(\frac{\tau}{2})\eta(\frac{\tau}{3})\eta(\frac{\tau}{5})\eta(\frac{\tau}{30})}\right)^{24}.
\end{equation}

\section{$G$ functions of $E_6$ and $E_7$}
\setcounter{equation}{0}

\subsection{Functions of $E_6$}
We introduce $G$ functions given by $SL(2,{\bf Z})$ transformation from $G_0(\tau)$. Furthermore we add some $G$ functions, so that all $G$ functions satisfy
the modular properties of $SU(3)$.
\beq
G_j(\tau):=\left(
\frac{\eta(\frac{\tau+j}{4})\eta(\frac{\tau+j}{6})}{\eta(\frac{\tau+j}{2})\eta(\frac{\tau+j}{3})\eta(\frac{\tau+j}{12})}
\right)^{24},j=0,...,11,
\enq
\beq
H_j(\tau):=\left(
\frac{\eta(\frac{2\tau+j}{2})\eta(\frac{2\tau+j}{3})}{\eta(2\tau)\eta(\frac{\tau-j}{3})\eta(\frac{2\tau+j}{6})}
\right)^{24},j=0,...,6,
\enq
\beq
I_j(\tau):=\left(
\frac{\eta(\frac{\tau-j}{4})\eta(\frac{3\tau+j}{2})}{\eta(\frac{\tau+j}{2})\eta(3\tau)\eta(\frac{3\tau+j}{4})}
\right)^{24},j=0,...,3,
\enq
\beq
J_j(\tau):=\left(
\frac{\eta(4\tau)\eta(\frac{2\tau+j}{3})}{\eta(2\tau)\eta(\frac{\tau-j}{3})\eta(\frac{4\tau+2j}{3})}
\right)^{24},j=0,...,2,
\enq
\beq
K_j(\tau):=\left(
\frac{\eta(\frac{2\tau+j}{2})\eta(6\tau)}{\eta(2\tau)\eta(3\tau)\eta(\frac{6\tau+j}{2})}
\right)^{24},j=0,1,
\enq
\beq
L_0(\tau):=\left(
\frac{\eta(4\tau)\eta(6\tau)}{\eta(2\tau)\eta(3\tau)\eta(12\tau)}
\right)^{24},
\enq
\beq
M_j(\tau):=\left(
\frac{\eta(\tau)\eta(\frac{3\tau+3j}{2})}{\eta(\frac{\tau+j}{2})\eta(3\tau)\eta(3\tau)}
\right)^{24},j=0,1,
\enq
\beq
N_j(\tau):=\left(
\frac{\eta(\tau)\eta(\frac{\tau+j}{6})}{\eta(\frac{\tau+j}{2})\eta(\frac{\tau+j}{3})\eta(\frac{\tau+j}{3})}
\right)^{24},j=0,...,5.
\enq
To obtain $G_j(\tau),\ldots, L_0(\tau)$, we have to take account of the 
structure of order $12$ Hecke transformation. 
$M_j(\tau)$ and $N_j(\tau)$ are obtained 
from $SL(2,{\bf Z})$ transformation of these functions. 
\subsection{Function of $E_7$}

\beq
G_j(\tau):=\left(
\frac{\eta(\frac{\tau+j}{6})\eta(\frac{\tau+j}{9})}{\eta(\frac{\tau+j}{2})\eta(\frac{\tau+j}{3})\eta(\frac{\tau+j}{18})}
\right)^{24},j=0,...,17,
\enq
\beq
H_j(\tau):=\left(
\frac{\eta(\frac{2\tau+2j}{3})\eta(\frac{\tau+j}{9})}{\eta(2\tau)\eta(\frac{\tau+j}{3})\eta(\frac{2\tau+2j}{9})}
\right)^{24},j=0,...,8,
\enq
\beq
I_j(\tau):=\left(
\frac{\eta(\frac{3\tau+j}{2})\eta(\frac{3\tau+j}{3})}{\eta(\frac{\tau+j}{2})\eta(3\tau)\eta(\frac{3\tau+j}{6})}
\right)^{24},j=0,...,5,
\enq
\beq
J_j(\tau):=\left(
\frac{\eta(6\tau)\eta(\frac{3\tau+j}{3})}{\eta(2\tau)\eta(3\tau)\eta(\frac{6\tau+2j}{3})}
\right)^{24},j=0,...,2,
\enq
\beq
K_j(\tau):=\left(
\frac{\eta(\frac{3\tau+j}{2})\eta(9\tau)}{\eta(\frac{\tau+j}{2})\eta(3\tau)\eta(\frac{9\tau+j}{2})}
\right)^{24},j=0,1,
\enq
\beq
L_0(\tau):=\left(
\frac{\eta(6\tau)\eta(9\tau)}{\eta(2\tau)\eta(3\tau)\eta(18\tau)}
\right)^{24},
\enq
\beq
M_j(\tau):=\left(
\frac{\eta(\tau)\eta(\frac{2\tau+2j}{3})}{\eta(2\tau)\eta(2\tau)\eta(\frac{\tau+j}{3})}
\right)^{24},j=0,1,2,
\enq
\beq
N_j(\tau):=\left(
\frac{\eta(\tau)\eta(\frac{\tau+j}{6})}{\eta(\frac{\tau+j}{2})\eta(\frac{\tau+j}{2})\eta(\frac{\tau+j}{3})}
\right)^{24},j=0,...,5.
\enq

\end{document}